# Realization of the Trajectory Propagation in the MM-SQC Dynamics by Using Machine Learning

Kunni Lin[1,2,‡], Jiawei Peng[1,2,‡], Chao Xu[2,3], Feng Long Gu[2,3,*] and Zhenggang Lan[2,3,*]

*1 School of Chemistry, South China Normal University, Guangzhou 510006, P. R. China.*

*2 MOE Key Laboratory of Environmental Theoretical Chemistry, South China Normal University, Guangzhou 510006, P. R. China.*

*3 SCNU Environmental Research Institute, Guangdong Provincial Key Laboratory of Chemical Pollution and Environmental Safety, School of Environment, South China Normal University, Guangzhou 510006, P. R. China.*

‡ *Co Contributors.*

\* *Corresponding Author. E-mail: gu@scnu.edu.cn; zhenggang.lan@m.scnu.edu.cn.*

**Abstract**

The supervised machine learning (ML) approach is applied to realize the trajectory-based nonadiabatic dynamics within the framework of the symmetrical quasi-classical dynamics method based on the Meyer-Miller mapping Hamiltonian (MM-SQC). After the construction of the long short-term memory recurrent neural network (LSTM-RNN) model, it is used to perform the entire trajectory evolutions from initial sampling conditions. The proposed idea is proven to be reliable and accurate in the simulations of the dynamics of several site-exciton electron-phonon coupling models, which cover two-site and three-site systems with biased and unbiased energy levels, as well as include a few or many phonon modes. The LSTM-RNN approach also shows the powerful ability to obtain the accurate and stable results for the long-time evolutions. It indicates that the LSTM-RNN model perfectly captures of dynamical correction information in the trajectory evolution in the MM-SQC dynamics. Our work provides the possibility to employ the ML methods in the simulation of the trajectory-based nonadiabatic dynamic of complex systems with a large number of degrees of freedoms.

With the rapid development of computer facilities and artificial intelligence sciences in recent years, machine learning (ML) methods are extensively used in theoretical simulations of chemical dynamics, such as the construction of the potential energy surfaces,[1-5] the automatic analysis of molecular dynamics evolution[6-8] and the numerical solution of the dynamical equation.[9-24] Among these works, considerable efforts were made to directly employ the various ML approaches to propagate the nonadiabatic dynamics evolutions,[9, 12-17, 20-21, 23] in which the strong electronic-nuclear couplings exist.

Nonadiabatic dynamics plays an essential role in the photoinduced molecular processes.[25-30] Therefore, many dynamical approaches were developed to describe nonadiabatic transitions.[27-28, 30-41] In recent years, several works tried to introduce ML approaches to realize the long-time nonadiabatic dynamics propagation by learning the short-time one.[9, 11-12, 14-15, 23] Some of these works are relevant to an interesting idea of the transfer tensor[42]. Under the assumption of the time-translational invariance, this approach builds the transfer tensor model to capture the essential dynamics features by learning the short-time evolution, and this transfer tensor can be used to propagate the long-time dynamics. Such idea was also realized to propagate the reduced quantum dynamics for a given system-plus-bath model by using different machine learning models, such as different neural network approaches.[9, 11-12, 14-15, 23] These works showed the possibilities to apply the ML methods to deal with the nonadiabatic dynamics propagation. In addition, it is also possible to build a unified ML model for a group of system-plus-bath Hamiltonians with different parameters.[13, 16] Many of these works more-or-less require the early short-time dynamics data in the ML model construction in order to make the long-time dynamics prediction.

Alternatively, it is also possible to employ the ML model to achieve the open quantum dynamics simulation only based on initial conditions, instead of a piece of the prior short-time dynamical evolution.[10, 20-21] For instance, Banchi et. al[10] combined the recurrent neural network (RNN) with gated recurrent unit (GRU) and quantum master equation to model the non-Markovian quantum processes starting from different initial states. Ullah et. al[20-21] employed the convolutional neural network (CNN) approach to achieve the prediction of the excitation energy transfer in the light-harvesting complexes. In the work purposed by Secor et. al,[43] the propagators of the time-dependent Schrödinger equation were "learned" by the artificial neural networks (ANN). Such ANN-based propagators can be used to simulate the quantum dynamics of systems efficiently for the given initial wavepackets. These works show the great potential of the ML application in nonadiabatic quantum dynamics. Besides, the ML approach was also used to propagate the classical dynamics of molecular systems,[22] when initial conditions at time zero were given.

In the simulation of the nonadiabatic dynamics, trajectory-based dynamics methods received the great attentions,[36-40, 44-55] because they can be used to treat high-dimensional systems with complicated molecular motions at all atomic levels. Among them, the symmetrical quasi-classical dynamics method based on the Meyer-Miller mapping Hamiltonian (MM-SQC)[39, 56] was proven to be a promising method that gives the proper description of the nonadiabatic dynamics in both model and the realistic molecular systems.[57-59] In the MM-SQC method, the trajectory propagation not only provides the descriptions on the electronic motions, but also gives the detailed information of the evolution of all nuclear or vibrational degrees of freedoms (DOFs). It should be highly interesting to examine the possibility to use the ML model to predict the trajectory propagation in the MM-SQC dynamics and describe the evolution of all DOFs in the nonadiabatic dynamics.

For this purpose, we propose to use "one-to-many" long short-term memory recurrent neural network (LSTM-RNN) framework[60-61] to perform the trajectory propagation in the MM-SQC dynamics, as the LSTM-RNN models are confirmed as the reliable tools to treat time-series problems in various research fields.[62-64] First, a few of trajectories are calculated by the MM-SQC method, which are given as the input data to build the LSTM-RNN model. Next the constructed ML model is used to simulate the trajectory propagation of all DOFs (both electronic mapping variables and nuclear DOFs), starting from new initial sampling conditions. After the employing the window approach[39, 65] for the quantum assignment of final quantum state, the LSTM-RNN method provides the excellent dynamics simulation results with respect to the MM-SQC dynamics, *i.e.* shows the accurate prediction of not only the population transfer dynamics but also the evolution of all electronic and nuclear DOFs. Such ML approach is accurate and trustful to realize the entire trajectory evolution in the MM-SQC dynamics. This work provides the important idea that sheds light on the employment of the ML approaches in the trajectory-based nonadiabatic dynamics simulation of complex systems.

Theoretically, the Meyer-Miller (MM) mapping Hamiltonian constructs a mapping transformation from a set of discrete quantum states to a set of coupled harmonic oscillators,[46, 55] that is

$$H_{MM}(Q^N, P^N, x^e, p^e) = \sum_k \left[ \frac{1}{2}((x_k^e)^2 + (p_k^e)^2) - \gamma \right] H_{kk}(Q^N, P^N) + \frac{1}{2} \sum_{k \neq l} (x_k^e x_l^e + p_k^e p_l^e) H_{kl}(Q^N, P^N).$$

In the current electron-phonon coupled systems, the $(Q^N, P^N)$ denote the coordinates and momenta of vibrational DOFs, and the $(x^e, p^e)$ represent the coordinates and momenta of electronic mapping variables, respectively. $H_{kk}$ and $H_{kl}$ denote the diagonal and off-diagonal elements of the Hamiltonian matrix, respectively. And the $\gamma$ is the parameter for the zero-point energy correction. This mapping Hamiltonian was initially proposed by Meyer and Miller,[46] and was lately re-visited formally by Stock and Thoss.[55] When all quantum operators are replaced by their classical correspondences, the classical MM mapping Hamiltonian is obtained, which can be used to run the classical dynamics. When both the initial sampling and the final assignment of the quantum states are performed within the window approach symmetrically, the MM-SQC method is constructed. The triangle window function[65] (with $\gamma = 1/3$) proposed by Cotton and Miller was employed in the MM-SQC dynamics. More theoretical details of the MM-SQC dynamics are given in the Supporting Information (SI). After the dynamics simulation, the final time-dependent populations of electronic states are given by averaging the binned quantum state assignment over all trajectories.

The site-exciton electron-phonon coupling models are widely used to describe the excited state energy transfer, in which each site describes a local excited state and the phonon modes are described by the harmonic oscillators. In the current work, we considered six site-exciton models, Model I-VI (see SI for details). For all models, each local excited state couples with its own vibrational DOFs. Model I and II denote the symmetrical and unsymmetrical two-site models with the unbiased ($V_{11}-V_{22}=0$ eV and $V_{12}=0.2$ eV) and biased ($V_{11}-V_{22}=0.2$ eV and $V_{12}=0.2$ eV) site energies, respectively. Model III and IV define the three-site models. Here all three sites show the same electronic energy in Model III ($V_{11}-V_{22}=V_{11}-V_{33}=0$ eV and $V_{12}=0.2$ eV). In Model IV, all sites show different electronic energies, ($V_{11}-V_{22}=V_{11}-V_{33}=0.1$ eV and $V_{12}=0.2$ eV). For the above four models, each localized electronic state couples with its own set of phonon modes (only eight modes) with the electron-phonon coupling parameters given in Table S1 in SI. In fact, these electron-phonon coupling parameters are taken from previous studies on the excited-state energy transfer in the stacked PBI system.[66] These early works demonstrated that these modes are strong coupled with the local excited state, as the influences of other modes are rather minor. The last two models (Model V and VI) denote the symmetric and unsymmetrical two-site models with the unbiased ($V_{11}-V_{22}=0$ eV and $V_{12}=0.03$ eV) and biased ($V_{11}-V_{22}=0.03$ eV and $V_{12}=0.03$ eV) site energies, respectively, while each site couples to its own continuous bath with the electron-phonon coupling fully characterized the Debye spectral density with the characteristic frequency ($\omega_c=500$ cm$^{-1}$) and the reorganization energy ($\lambda=62.5$ cm$^{-1}$). Since the MM-SQC dynamics requires the explicit description of the bath modes, we use 70 discretized modes (see SI) to represent each bath and therefore we totally have 140 phonon modes.

Here the initial condition of the MM-SQC dynamics is defined as the vertical excitation of the lowest vibrational level of the electronic ground state to the first local excited state. The samplings of the electronic mapping variables were conducted with the combination of the window trick and the action-angle approach, while the samplings of different phonon modes were performed by using the action-angle approach.

The working process of the dynamics simulation employed LSTM-RNN in our current work is shown in Scheme 1. The implementation details are given as follows.

We first generated 500 trajectories in the MM-SQC dynamics propagation up to 100 fs (for Model I-IV) or 200fs (for Model V-VI) to construct the input dataset for each individual ML model. For the *k-th* trajectory, all coordinates and momenta of nuclear DOFs and electronic mapping variables at a single time step constitute a vector $X^{(k)}(T)$ with the dimension $2N_v+2N_e$, in which the $N_v$ and $N_e$ represents the numbers of nuclear and electronic degrees. Taking the evolution of the Model I-IV as the example, the trajectory propagation can be represented as a time-series $[X^{(k)}(T_0), X^{(k)}(T_1), \cdots, X^{(k)}(T_{100})]$ with the discretized time step of 1 fs. With the defined length *L*, the time-series in each trajectory is split into (100-*L*+2) small time sequences $S_i^{(k)}$ according to the chronological order as shown as Scheme 1(b). These data subsets $\{S_i^{(k)}\}$ were randomly divided into two groups $\{S_i^{(k,A)}\}$ and $\{S_i^{(k,B)}\}$ in the ratio of 3:1. When all trajectories for a site-exciton model were processed, all $\{S_i^{(k,A)}\}$ were merged as the training dataset $\{S^{(A)}\}$ and all $\{S_i^{(k,B)}\}$ were merged as the validation dataset $\{S^{(B)}\}$. After the randomly shuffled operation with the individual chronological order unchanged, these two groups (training and validation datasets) were taken as the input data to construct the LSTM-RNN model.

Scheme 1(a) shows the structure of the LSTM-RNN model used in the current work. The NN model is consisted of the input layer, the LSTM layer, the dense layer and the output layer. For each time sequence, only the first vector in this series is passed into the LSTM layer as the input data. Then for the subsequent time steps, the LSTM layer only takes the previous LSTM output as the input data. This process was achieved by the gate mechanisms in the LSTM cells and the detailed explanations of the LSTM structures are discussed in previous works.[9, 12, 23, 60-61] After the analysis and transmission of effective information, the dense layer merges the results of the LSTM layer and then gives the whole time sequences. This "one-to-many" LSTM-RNN structure well captures the time-correlated information among several successive dynamical steps, which should be critical in the dynamical evolution.[42] In principle, it is difficult to extract the whole evolution dynamics only from the initial conditions, because in principle the correlations of the dynamical variables at the time zero and the later time step should become smaller with time being, particular for the nonlinear dynamics. Therefore, it is suitable to split the whole time series into many short-time sequences to improve the performance of the LSTM-RNN model.

The detail parameters of the constructions of these LSTM-RNN models are shown in the SI. After the training process, the reasonable LSTM-RNN models were obtained with the small validation loss. Next, we preformed the initial sampling again to generate the 2500 new sampling conditions. Starting from them, we employed the built LSTM-RNN model to predict the evolutions of all DOFs, including the coordinates and momenta of both nuclear DOFs and electronic mapping variables. Based on the initial dynamical-variable vector $X(T_0)$ as input, the LSTM-RNN model gives the values of $[X(T_1), \cdots X(T_{L-1})]$. And then, the last forecasted vector $X(T_{L-1})$ was used as the new input of the LSTM-RNN model to predict the $[X(T_L), \cdots X(T_{2L-2})]$. In this way, the entire evolution of 2500 new trajectories were obtained step by step by the built LSTM-RNN model up to 100 fs (for Model I-IV) or 200 fs (for Model V-VI). After all trajectories are obtained, we calculated the occupation of the quantum states along with the time evolution by the triangle window function, giving the electronic population dynamics.

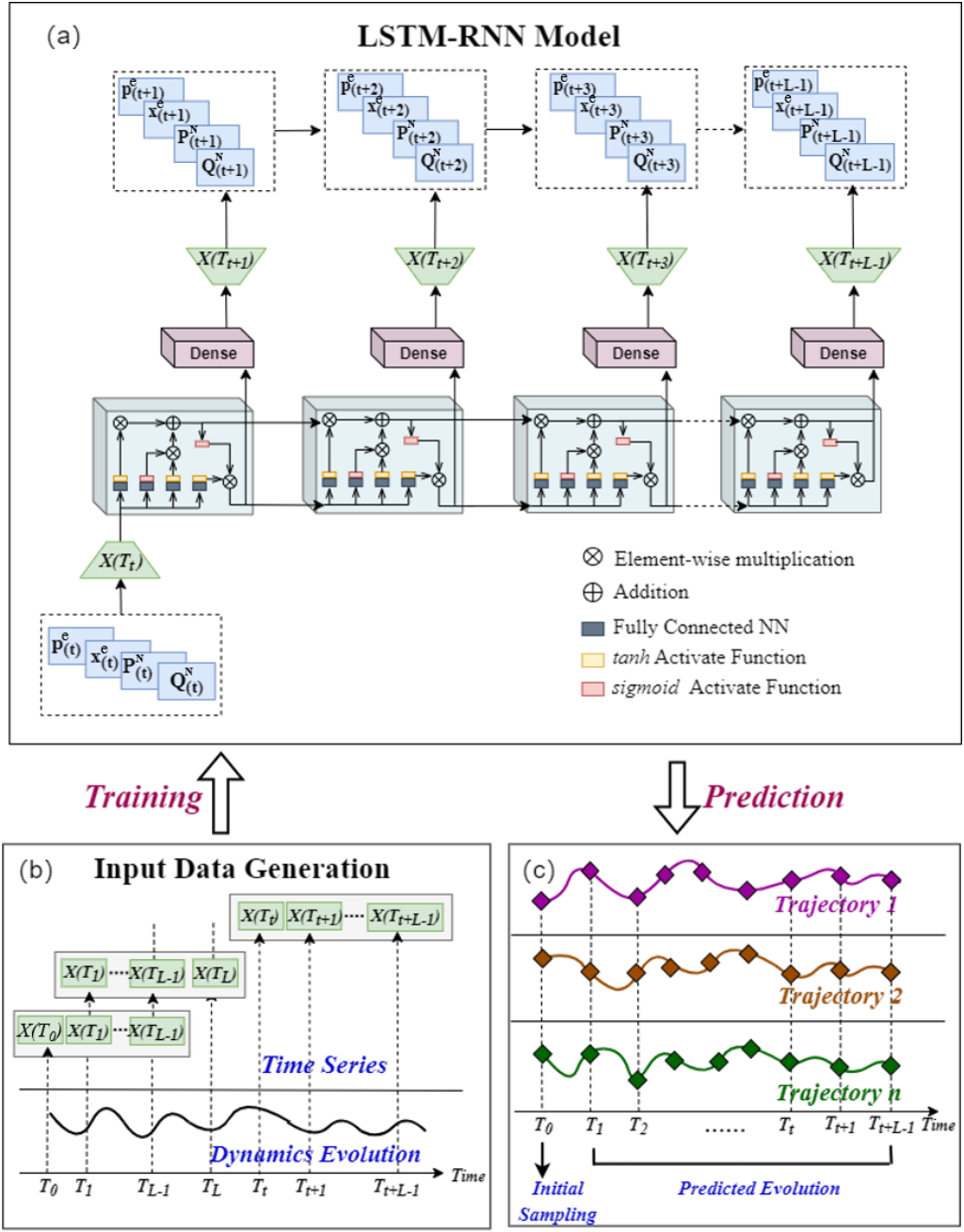

Scheme 1. The diagram of the LSTM-RNN model construction (a); the generation of the time series as input data (b) and the prediction of the entire evolution of the new dynamics trajectories (c).

Following the above working procedure, we employed the LSTM-RNN model to simulate trajectory propagation of Model I-VI, starting from 2500 new sampled initial conditions. The population dynamics given by the LSTM-RNN prediction agree very well with the simulation results by the MM-SQC method (shown in Figure 1), no matter whether the models include two or three sites with the symmetric or asymmetric site energies. Even for the Model V and Model VI that contain a huge number of DOFs, the prediction results of the population dynamics are still reasonable. In all Models, the LSTM-RNN method well reproduces all features in the MM-SQC dynamics, including the population oscillation period and amplitude, as well as all decay patterns. These results indicate that the LSTM-RNN approach is a powerful tool to perform the MM-SQC dynamics.

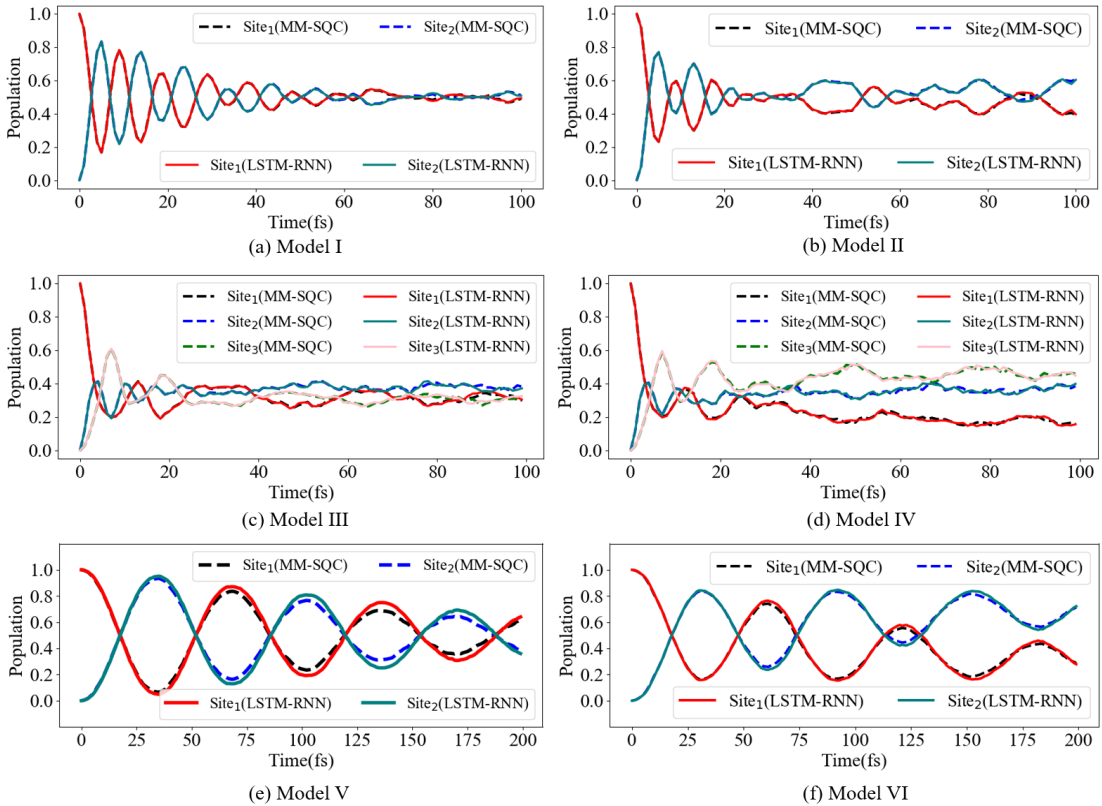

Figure 1. The time-dependent electronic populations given by the LSTM-RNN prediction *vs.* the MM-SQC dynamics simulation of Model I (a), Model II (b), Model III (c), Model IV (d),

Model V (e) and Model VI (f) respectively. Among that, the evolution time of the dynamics of Model I-IV and Model V-VI are 100 fs and 200 fs, respectively.

Besides the overall dynamics population, it is also important to know whether the LSTM-RNN model can give the reasonable evolution of all individual dynamical variables, including the electronic mapping variables $(x^e, p^e)$ and all vibrational DOFs $(Q^N, P^N)$. For this purpose, we take the Model I as a prototype to analyse the LSTM-RNN dynamics in details.

In order to compare the evolution of the electronic mapping variables over the 2500 new trajectories by the LTSM-RNN prediction and the MM-SQC dynamics, we show their time-dependent coordinate distributions (see Figure 2). The coordinate distributions with time being in these two simulation approaches are highly consistent, which give exactly the same peak features such as their time-space locations, intensities and shapes. We also noticed that the results of both approaches display the same recurrence phenomenon along with time being. For the momenta of electronic mapping variables ($p_1^e$ and $p_2^e$), the LSTM-RNN method also gives the excellent results as shown in the Figure S1 in SI. To obtain the expressiveness of the LTSM-RNN model in a single trajectory, we randomly selected 100 from 2500 trajectories and displayed the time evolution of $(x^e, p^e)$ by the LSTM-NN method along with the MM-SQC simulation results as standard data in Figure S2-S3 in the SI. All forecasted variables are consistent with the MM-SQC evolution.

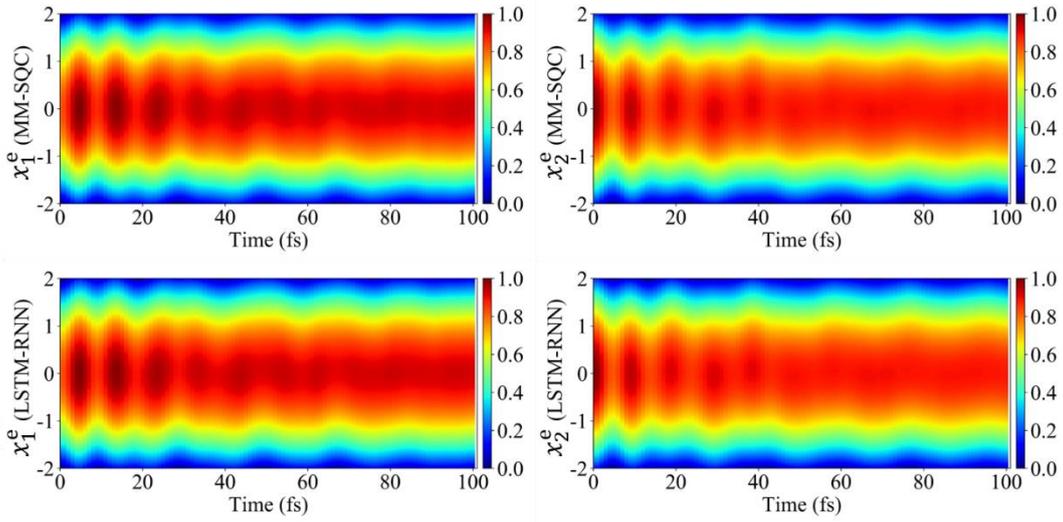

Figure 2. The comparison of the coordinate distributions of two sites between the LSTM-RNN prediction and the MM-SQC simulation of Model I. The $x_1^e$ represents the coordinate of the first site and the $x_2^e$ represents the another one.

For all nuclear DOFs $(Q^N, P^N)$, we calculated their prediction errors at different time steps, which are described by the mean absolute errors (MAEs) between the LSTM-RNN prediction values and the MM-SQC simulation results over total 2500 new trajectories. We selected a few discrete time steps (20 fs, 40 fs, 60 fs, 80 fs, 100 fs) to show the MAEs of all pairs of $(Q^N, P^N)$ in Figure 3. It is clearly that the LSTM-RNN model provides the accurate description of the evolution of the nuclear DOFs. Within the dynamics evolution of the first 60 fs, the MAEs of all vibrational modes remain very small. When the dynamics is propagated up to 100 fs, the deviations between the LSTM-RNN model and MM-SQC method appear for some modes. This indicates that the prediction errors increase rather slowly with time being, possibly due to the error accumulation in the prediction process.

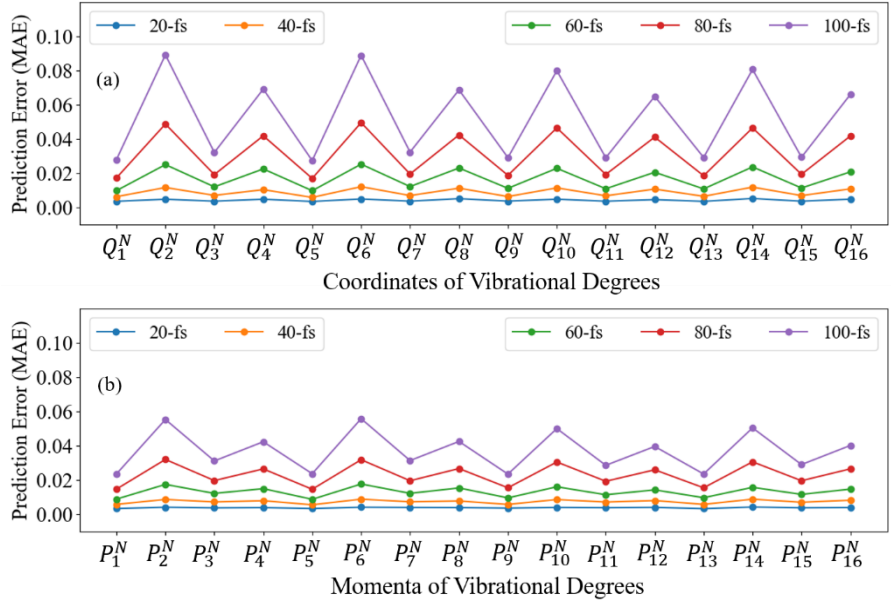

Figure 3. The prediction errors of all vibrational degrees at 20 fs, 40 fs, 60 fs, 80 fs, 100 fs of Model I, which defined by the MAEs of 2500 new trajectories in corresponding time steps.

As a short summary, the LSTM-RNN method well reproduces the MM-SQC dynamics, which provides the reliable description of the electronic population dynamics and the evolution of all DOFs, even for the propagation of each individual trajectory.

In order to understand the extensibility of the current LSTM-RNN model, we also used the built networks to propagate the MM-SQC dynamics up to the longer time 400 fs. In Figure 4 the population dynamics in both LSTM-RNN predictions and the MM-SQC simulations are still consistent, even if the LSTM-NN model is built from the short-time dynamics data (100 fs for Model I-IV and 200 fs for Model V-VI). These results indicate that the MM-SQC dynamics evolution within the short time duration contains the enough features to characterize the long-time evolution. As such dynamical features were properly captured by the LSTM-RNN model, it can be used to predict not only the short-time dynamics evolution but also the long-time propagation results.

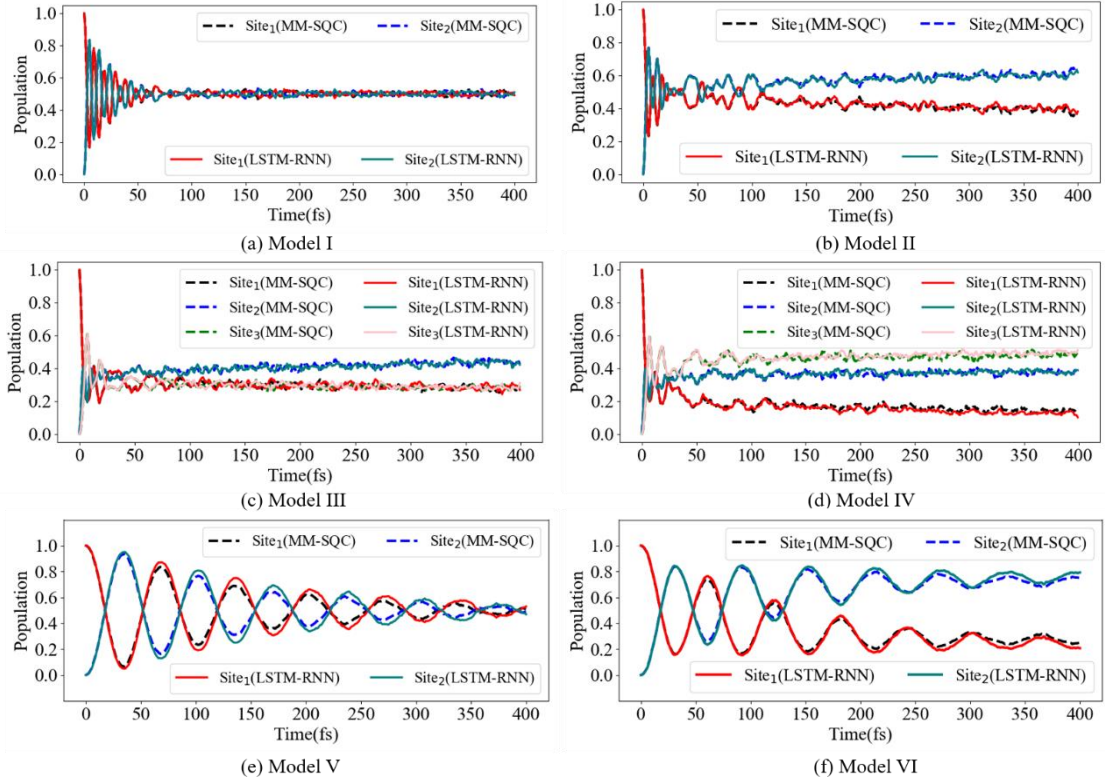

Figure 4. The time-dependent electronic populations within 400 fs given by the LSTM-RNN prediction *vs.* the MM-SQC dynamics simulation of Model I-VI.

## IV. Conclusion

The current work employs the LSTM-RNN approach to perform the MM-SQC dynamics simulations. After the construction of the LSTM-RNN model, we can employ it to realize the accurate trajectory evolution in the MM-SQC dynamics only starting from the initial condition. We use different site-exciton models to examined the performance of this approach, which cover two-site and three-site models with biased and unbiased energy levels, as well as include a limited number or a large number of vibrational DOFs. All LSTM-RNN predictions show the perfect results confirmed by the MM-SQC dynamics simulations. This indicates that the LSTM-RNN well captures the key evolution features of the trajectory propagation. Such excellent ability allows the LSTM-RNN models to realize the propagation of a large number of trajectories even with the longer time scales.

In conclusion, the LSTM-RNN is a powerful approach for constructing the reliable trajectory propagation at the same level as the MM-SQC method. This work provides a good foundation for performing the trajectory-based nonadiabatic dynamics simulation of complex systems. For more expectations, it is possible to employ the current LSTM-RNN approach to simulate the on-the-fly MM-SQC dynamics. This provides a unique idea to explore the nonadiabatic dynamics of realistic polyatomic systems at the atomic level by using the massive trajectory propagation with rather manageable computational cost.

**Supporting Information Available**

Several relevant information: The detailed construction of triangle window function and action variable in the MM-SQC dynamics; the detailed descriptions of site-exciton Hamiltonians; the parameters of all models; the construction of LSTM-RNN models; the comparison of the distributions of the electronic mapping momenta between the LSTM-RNN and MM-SQC results in Model I; and the evolution of the coordinates and momenta of vibrational DOFs in 100 trajectories randomly selected from 2500 new trajectories in Model I.

**Author Information**

**Corresponding Author**

E-mail: gu@scnu.edu.cn; zhenggang.lan@m.scnu.edu.cn.

**Notes**

The authors declare no competing financial interest.

**Acknowledgments**

The authors express sincere to thank the National Natural Science Foundation of China (No. 21873112, 21933011 and 21903030) for financial support. Some calculations in this paper were done on SunRising-1 computing environment in Supercomputing Center, Computer Network Information Center, CAS.

# Supporting Information for

# Realization of the Trajectory Propagation in the MM-SQC Dynamics by Using Machine Learning

Kunni Lin[1,2,‡], Jiawei Peng[1,2,‡], Chao Xu[2,3], Feng Long Gu[2,3,*] and Zhenggang Lan[2,3,*]

*1 School of Chemistry, South China Normal University, Guangzhou 510006, P. R. China.*

*2 MOE Key Laboratory of Environmental Theoretical Chemistry, South China Normal University, Guangzhou 510006, P. R. China.*

*3 SCNU Environmental Research Institute, Guangdong Provincial Key Laboratory of Chemical Pollution and Environmental Safety, School of Environment, South China Normal University, Guangzhou 510006, P. R. China.*

*‡ Co Contributors.*

*\* Corresponding Author. E-mail: gu@scnu.edu.cn; zhenggang.lan@m.scnu.edu.cn.*

## S1. *Triangle Window Function and Action Variable.*

The triangle window function was employed in both the initial-state sampling and the final product assignment in the MM-SQC dynamics simulation. Here, we give an example of such window function for a two-state model as

$$W_1(n_1, n_2) = 2 \cdot h(n_1 + \gamma - 1)h(n_2 + \gamma)h(2 - 2\gamma - n_1 - n_2),$$

$$W_2(n_1, n_2) = 2 \cdot h(n_1 + \gamma)h(n_2 + \gamma - 1)h(2 - 2\gamma - n_1 - n_2),$$

where $h(z)$ is the Heaviside function and the quantity $\gamma$ is defined as 1/3 according to previous works.

In the MM-SQC dynamics simulation, the action variable $n_k$ described as

$$n_k = \frac{1}{2}\left[(x_k)^2 + (p_k)^2\right] - \gamma$$

is used to assign the quantum level from the mapping variables. A large number of trajectories were run and the average of the binned quantum levels finally gives the time-dependent populations of electronic states.

## S2. Hamiltonian.

In currenrt work, the site-exciton electron-phonon coupling Hamiltonian was taken to define Model I-VI. In these models, the Hamiltonian is written as

$$H = H_e + H_{ph} + H_{e-ph}. \tag{1}$$

Here, the electronic system part written as

$$H_e = \sum_{k=1}^{n} |\varphi_k\rangle V_{kk} \langle\varphi_k| + \sum_{k \neq l} |\varphi_k\rangle V_{kl} \langle\varphi_k|, \tag{2}$$

is composed of $n$ local-excited (LE) electronic states. Each LE state couples with its own set of phonon modes. The phonon Hamiltonian is written as

$$H_{ph} = \sum_{k=1}^{n} \sum_{j}^{N_b} \frac{1}{2} \omega_{kj} (Q_{kj}^2 + P_{kj}^2). \tag{3}$$

And the bilinear electron-phonon interaction is considered as

$$H_{e-ph} = \sum_{k=1}^{n} |\varphi_k\rangle (\sum_{j}^{N_b} \kappa_{kj} Q_{kj}) \langle\varphi_k|. \tag{4}$$

In Eq.(2)-(4), $N_b$ represents the total number of phonon modes. $\omega_{kj}$, $Q_{kj}$, and $P_{kj}$, are the corresponding frequency, position and momentum of each phonon mode, respectively. The $\kappa_{kj}$ characterizes electron-phonon coupling strength. The subscripts $k$ and $j$ refer to the indices of the electronic state and the phonon mode, respectively.

*S3. Models.*

In this work, six electron-phonon coupling models are considered.

- Model I: symmetrical site-exciton model with the unbiased ($V_{11} - V_{22} = 0$ eV, $V_{12} = 0.2$ eV) site energy. The Hamiltonian is given as

$$\text{Model I:} \quad H = \begin{bmatrix} 0.0 & 0.2 \\ 0.2 & 0.0 \end{bmatrix}$$

- Model II: Asymmetric site-exciton model with the biased ($V_{11} - V_{22} = 0.2$ eV, $V_{12} = 0.2$ eV) site energy. The Hamiltonian is given as

$$\text{Model II:} \quad H = \begin{bmatrix} 0.2 & 0.2 \\ 0.2 & 0.0 \end{bmatrix}$$

- Model III: Three-site model with the unbiased ($V_{11} - V_{22} = V_{11} - V_{33} = 0$ eV, $V_{12} = 0.2$ eV) site energy. The Hamiltonian is given as

$$\text{Model III:} \quad H = \begin{bmatrix} 0.0 & 0.2 & 0.0 \\ 0.2 & 0.0 & 0.2 \\ 0.0 & 0.2 & 0.0 \end{bmatrix}$$

- Model IV: Three-site model with the biased ($V_{11} - V_{22} = V_{11} - V_{33} = 0.1$ eV, $V_{12} = 0.2$ eV) site energy. The Hamiltonian is given as

$$\text{Model IV:} \quad H = \begin{bmatrix} 0.2 & 0.2 & 0.0 \\ 0.2 & 0.1 & 0.2 \\ 0.0 & 0.2 & 0.0 \end{bmatrix}$$

- Model V: Two-site model with unbiased site energy unbiased ($V_{11} - V_{22} = 0$ eV, $V_{12} = 0.03$ eV) and the Hamiltonian is given as

$$\text{Model V:} \quad H = \begin{bmatrix} 0.00 & 0.03 \\ 0.03 & 0.00 \end{bmatrix}$$

- Model VI: Two-site model with biased site energy unbiased ($V_{11} - V_{22} = 0.03$ eV, $V_{12} = 0.03$ eV) and the Hamiltonian is given as

$$\text{Model VI:} \quad H = \begin{bmatrix} 0.03 & 0.03 \\ 0.03 & 0.00 \end{bmatrix}$$

For Model I-IV, each localized electronic state couples with only eight modes and their electron-phonon couplings are given in Table S1. In fact, these electron-phonon coupling parameters are taken from previous studies on the excited-state energy transfer in the stacked PBI system. These early works demonstrated that these modes are strongly coupled with the local excited state, as the influence of other modes are rather minor.

In the last two models (Model V and VI), the electron-phonon coupling is fully characterized by the Debye spectral density, *i.e.*

$$J(\omega) = \frac{2\lambda \omega \omega_c}{\omega^2 + \omega_c^2}, \tag{5}$$

where the $\omega_c$ represents the characteristic frequency and $\lambda$ refers to the reorganization energy. The spectral density is represented by a series of discretized bath modes which as

$$J_k(\omega) = \frac{1}{2}\pi \sum_{i=1}^{N} \kappa_{ki}^2 \delta(\omega - \omega_{ki}). \tag{6}$$

Here, when the sampling interval $\Delta\omega$ is given, $\kappa_{ki}$ is evaluated by

$$\kappa_{ki} = (\frac{2}{\pi} J_k(\omega_{ki})\Delta\omega)^{1/2}, \qquad (7)$$

where $\omega_c$ =500 cm$^{-1}$ and $\lambda$ =62.5 cm$^{-1}$.

In Model V-VI, 70 phonon modes were employed to characterize the electron-phonon couplings, which are given in Table S2.

Table S1. The parameters of the electron-phonon couplings in Model I-IV.

| ω [eV] | κ [eV] |
|---|---|
| 0.0680 | -0.0266 |
| 0.0811 | -0.0194 |
| 0.1649 | -0.1120 |
| 0.1727 | -0.0720 |
| 0.1748 | 0.0378 |
| 0.1823 | 0.0383 |
| 0.1991 | 0.1101 |
| 0.2020 | 0.0642 |

Table S2. The parameters of the electron-phonon couplings in Model V-VI.

| ω [eV] | κ [eV] |
|---|---|
| 0.00148782 | 0.00059338 |
| 0.00297563 | 0.00083844 |
| 0.00446345 | 0.00102541 |
| 0.00595126 | 0.00118167 |
| 0.00743908 | 0.00131777 |
| 0.00892689 | 0.00143906 |
| 0.01041471 | 0.00154869 |
| 0.01190252 | 0.00164871 |
| 0.01339034 | 0.00174052 |
| 0.01487815 | 0.00182515 |
| 0.01636597 | 0.00190338 |
| 0.01785378 | 0.00197582 |
| 0.01934160 | 0.00204296 |
| 0.02082941 | 0.00210521 |
| 0.02231723 | 0.00216293 |
| 0.02380504 | 0.00221641 |
| 0.02529286 | 0.00226594 |
| 0.02678067 | 0.00231174 |
| 0.02826849 | 0.00235404 |
| 0.02975630 | 0.00239305 |
| 0.03124412 | 0.00242894 |
| 0.03273193 | 0.00246191 |
| 0.03421975 | 0.00249211 |
| 0.03570756 | 0.00251970 |
| 0.03719538 | 0.00254483 |
| 0.03868320 | 0.00256765 |
| 0.04017101 | 0.00258828 |
| 0.04165883 | 0.00260686 |
| 0.04314664 | 0.00262349 |
| 0.04463446 | 0.00263831 |
| 0.04612227 | 0.00265142 |
| 0.04761009 | 0.00266293 |
| 0.04909790 | 0.00267293 |
| 0.05058572 | 0.00268151 |
| 0.05207353 | 0.00268878 |
| 0.05356135 | 0.00269480 |
| 0.05504916 | 0.00269967 |
| 0.05653698 | 0.00270345 |
| 0.05802479 | 0.00270622 |
| 0.05951261 | 0.00270806 |
| 0.06100042 | 0.00270901 |
| 0.06248824 | 0.00270914 |
| 0.06397605 | 0.00270851 |
| 0.06546387 | 0.00270717 |
| 0.06695168 | 0.00270518 |
| 0.06843950 | 0.00270257 |
| 0.06992731 | 0.00269940 |
| 0.07141513 | 0.00269570 |
| 0.07290294 | 0.00269152 |
| 0.07439076 | 0.00268689 |
| 0.07587858 | 0.00268184 |
| 0.07736639 | 0.00267641 |
| 0.07885421 | 0.00267063 |
| 0.08034202 | 0.00266452 |
| 0.08182984 | 0.00265812 |
| 0.08331765 | 0.00265145 |
| 0.08480547 | 0.00264453 |
| 0.08629328 | 0.00263738 |
| 0.08778110 | 0.00263002 |
| 0.08926891 | 0.00262247 |
| 0.09075673 | 0.00261476 |
| 0.09224454 | 0.00260689 |
| 0.09373236 | 0.00259888 |
| 0.09522017 | 0.00259075 |
| 0.09670799 | 0.00258250 |
| 0.09819580 | 0.00257416 |
| 0.09968362 | 0.00256573 |
| 0.10117143 | 0.00255723 |
| 0.10265925 | 0.00254866 |
| 0.10414706 | 0.00254004 |

## S4. The Construction parameters of LSTM-RNN Models.

The training data set was used to train the LSTM-RNN model and the validation set was used to evaluate the performance of the LSTM-RNN model. In all models, only one LSTM-RNN layer is considered.

For Model I-IV, we trained the LSTM-RNN model for 2000 epoches with $10^{-5}$ as the learning rate, 50 as the batch size, 5 as the sequence length and 2000 as the number of neurons in the LSTM layer.

For Model V-VI, we trained the LSTM-RNN model for 2000 epoches with $10^{-5}$ as the learning rate, 50 as the batch size, 20 as the sequence length and 2000 as the number of neurons in the LSTM layer.

## S5. The Momenta Distributions of Model I Comparised Between LSTM-RNN and MM-SQC.

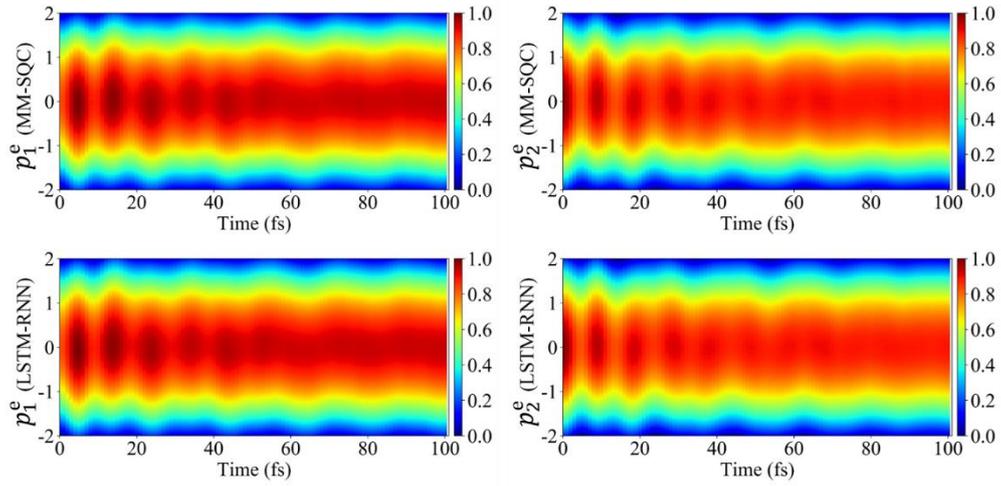

Figure S1. The comparison of the momentum distributions of electronic mapping variables between the LSTM-RNN prediction and the MM-SQC simulation results of Model I. The $p_1^e$ represents the momentum of the first LE site, and the $p_2^e$ represents the another one.

*S6. The Performance of the Coordinates and Momenta of Electronic Mapping Variables in Single Trajectories.*

Figure S2. The evolution of $(x_1^e, p_1^e)$ in 100 randomly selected trajectories calculated by the LSTM-RNN prediction *vs.* the MM-SQC dynamics simulation of Model I.

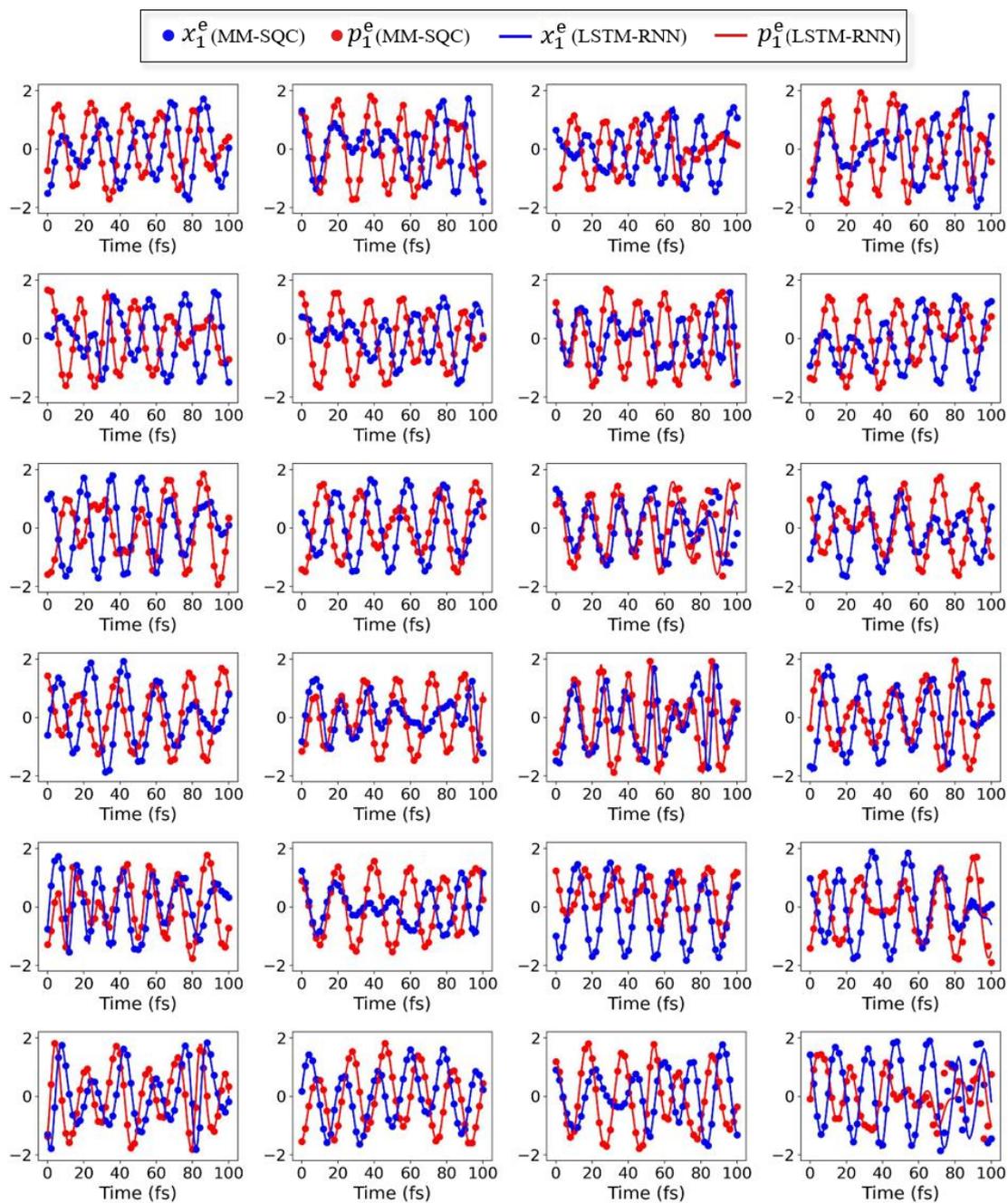

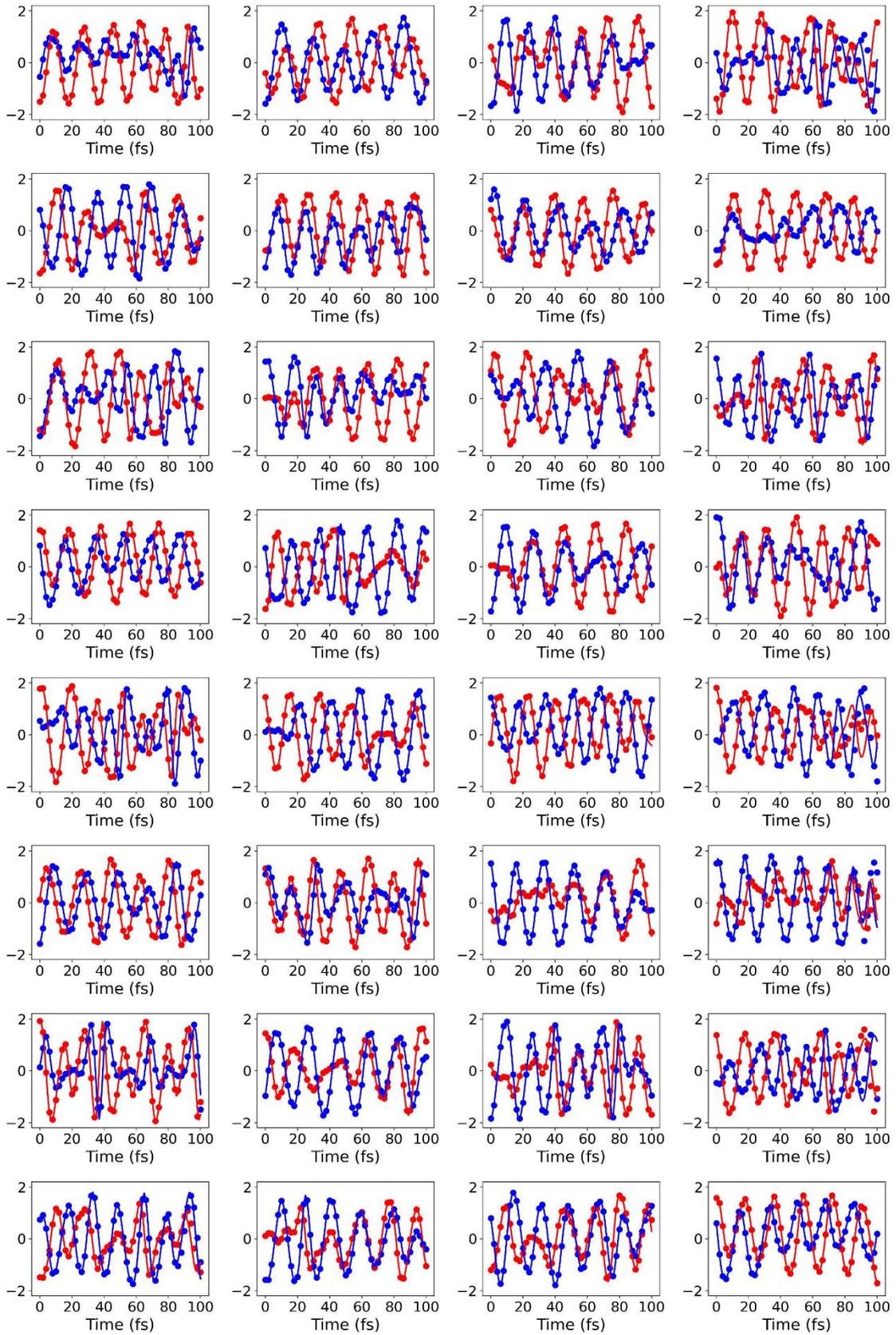

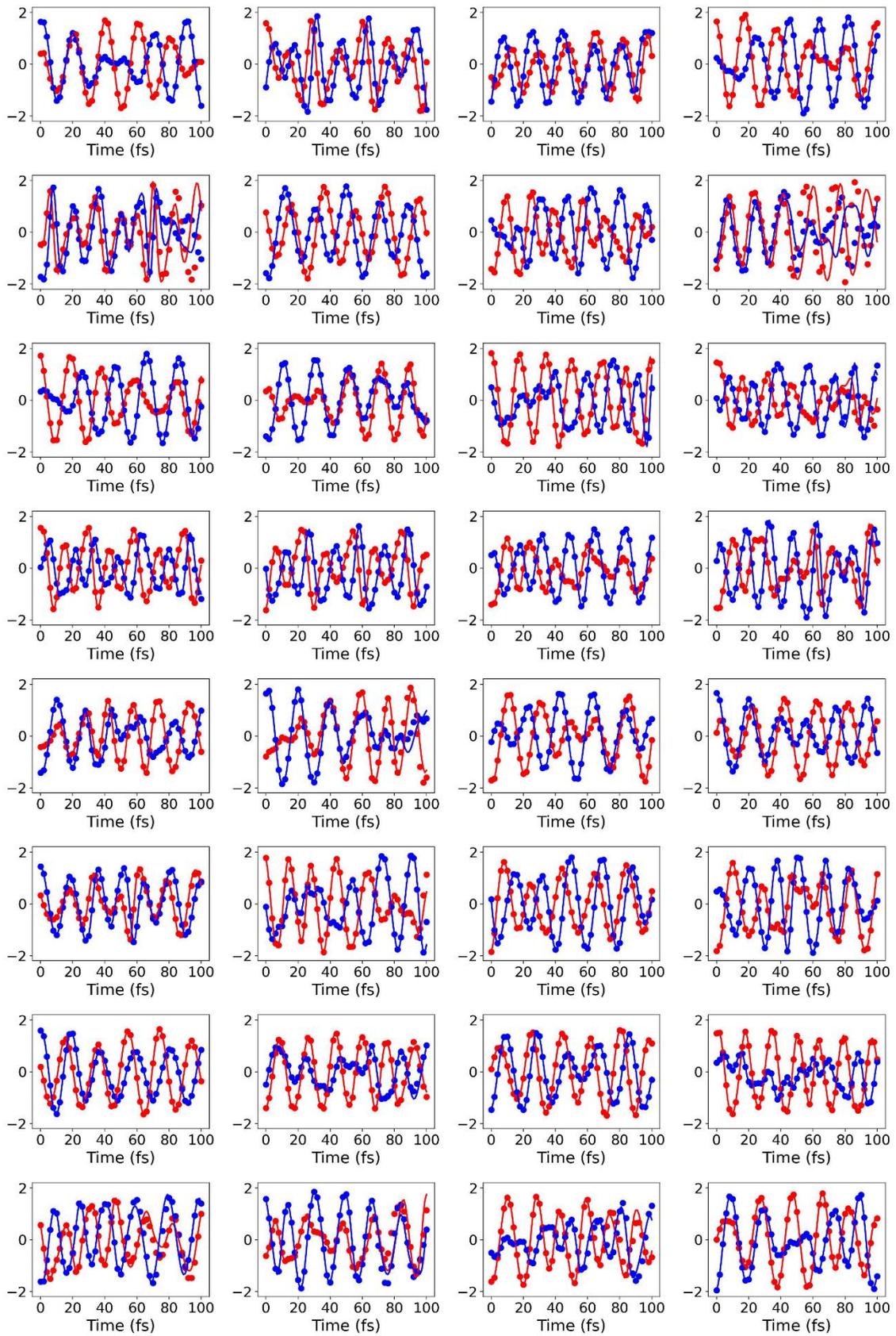

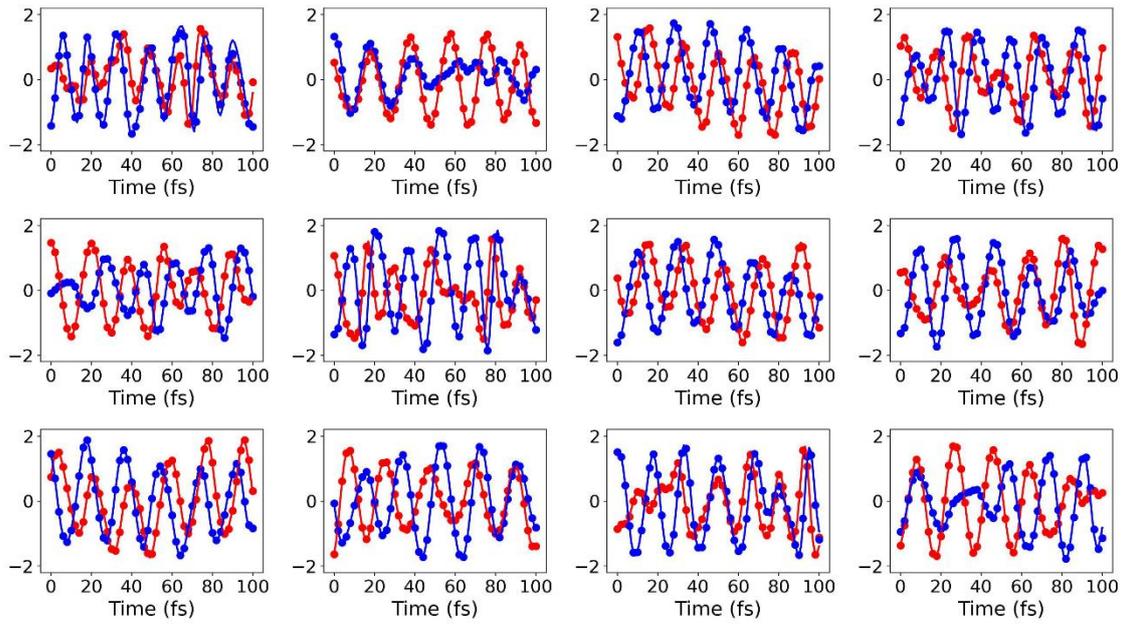

Figure S3. The evolution of $(x_2^e, p_2^e)$ in 100 randomly selected trajectories calculated by the LSTM-RNN prediction *vs.* the MM-SQC dynamics simulation of Model I.

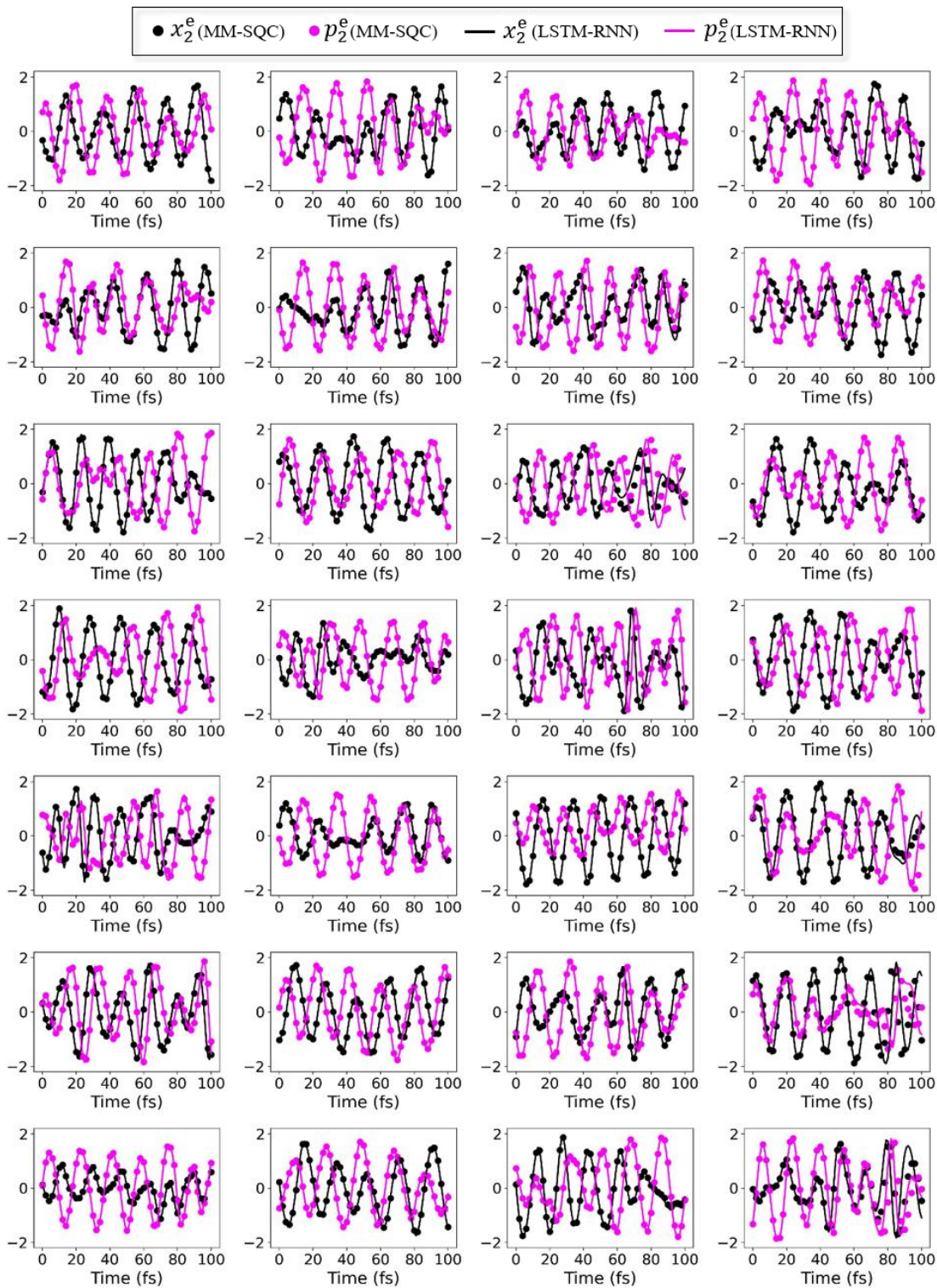

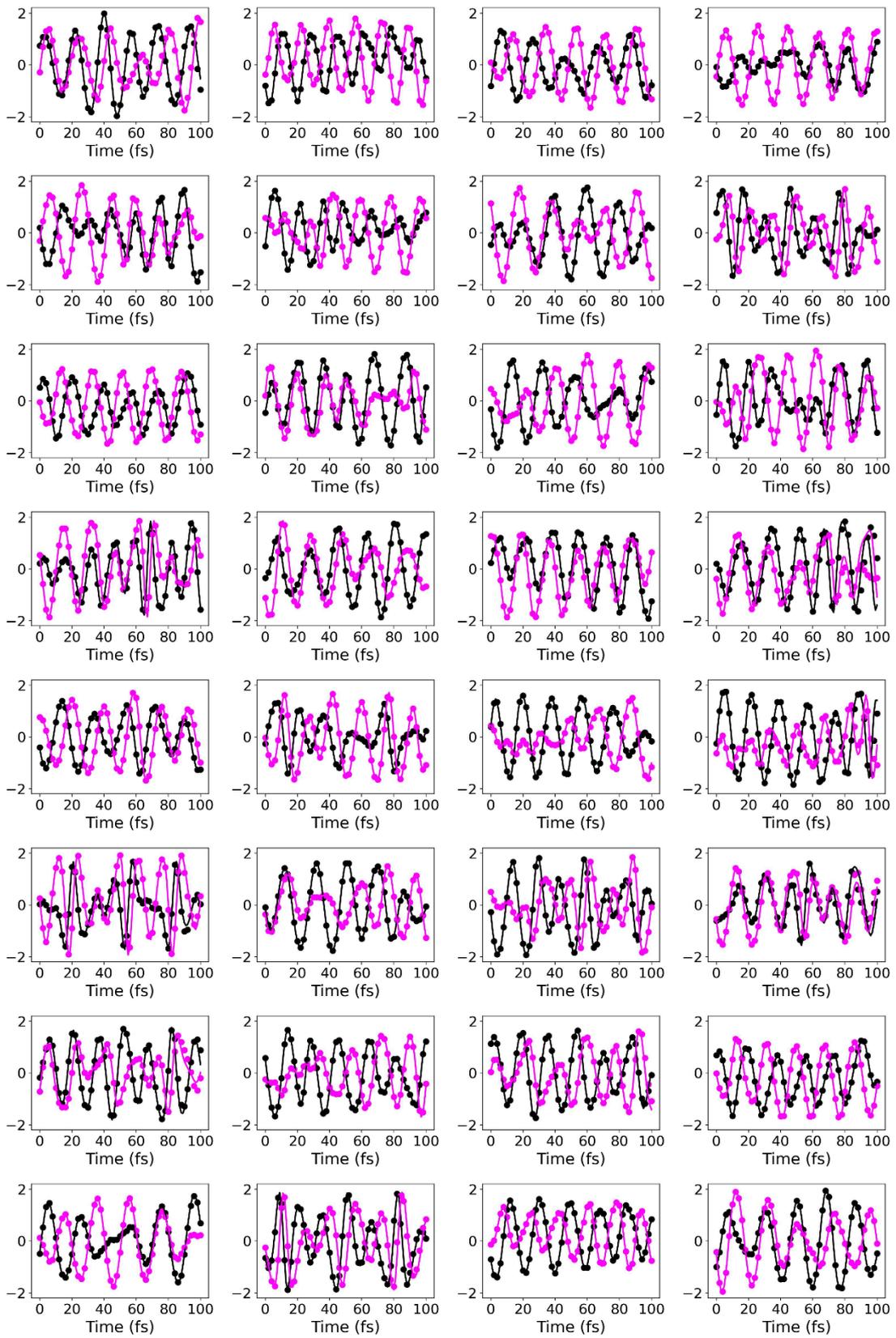

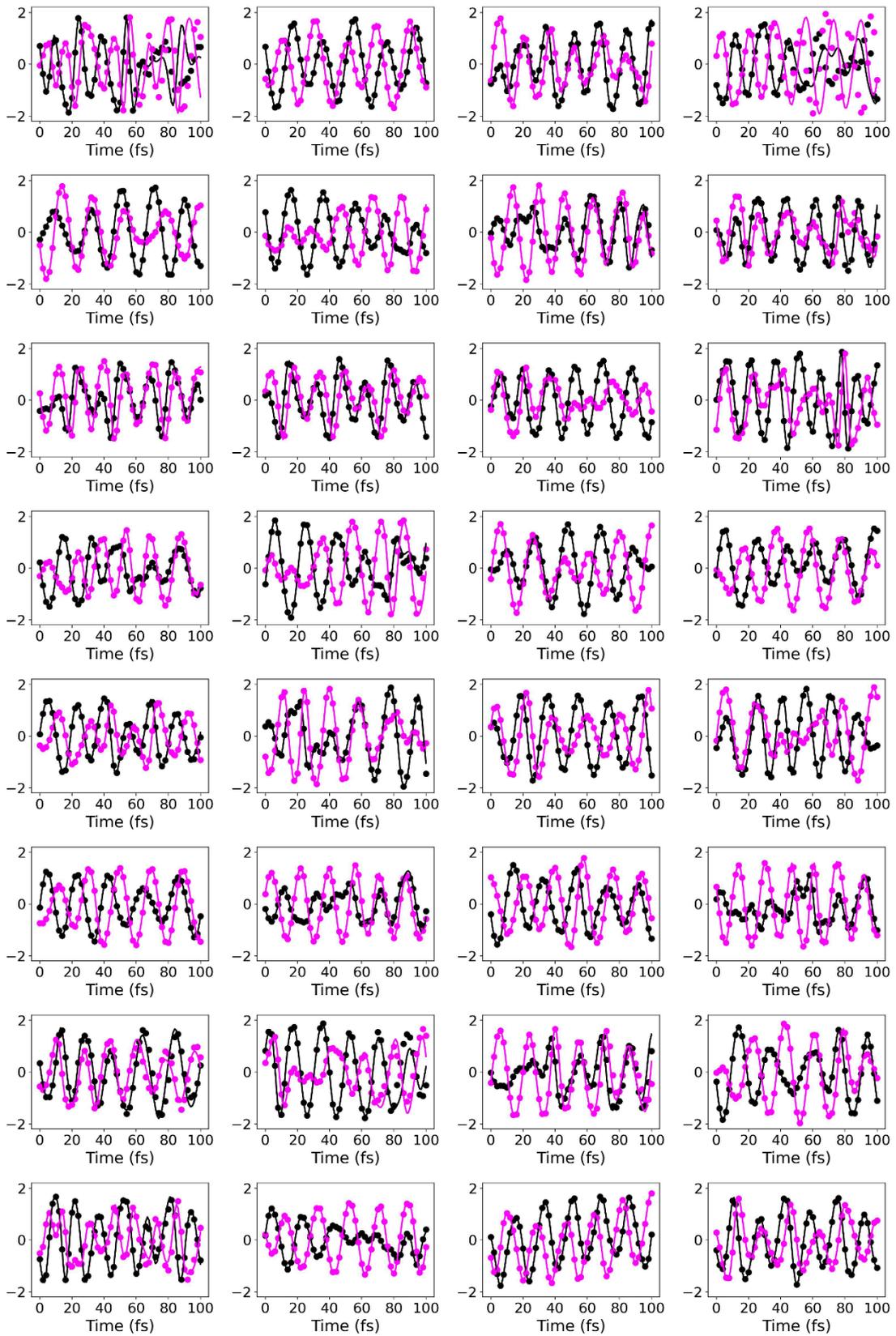
45

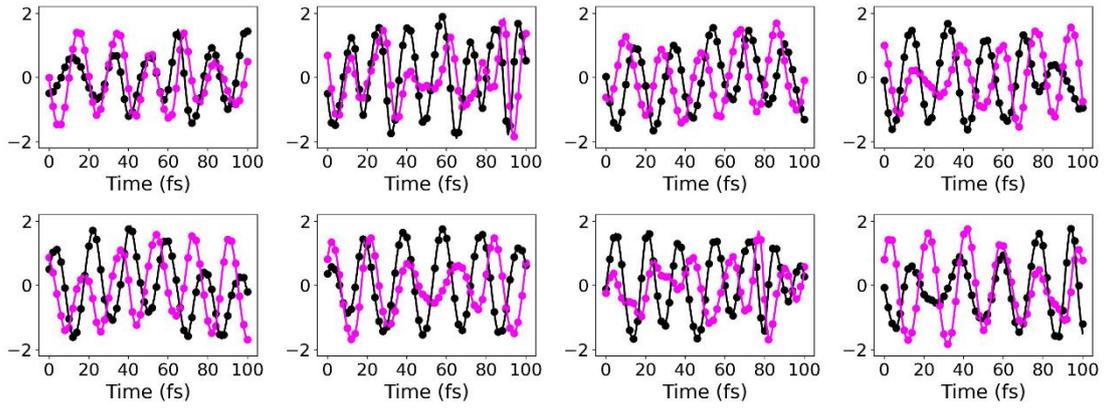